\documentclass{PoS}
\usepackage{amssymb,amsmath,epsfig}
\newcommand{\eq}[1]{Eq.~(\ref{#1})}

\newcommand{\be}{\begin{equation}}
\newcommand{\ee}{\end{equation}}
\newcommand{\bea}{\begin{eqnarray}}
\newcommand{\eea}{\end{eqnarray}}
\newcommand{\ben}{\begin{eqnarray*}}
\newcommand{\een}{\end{eqnarray*}}

\newcommand{\w}{\omega}
\newcommand{\e}{\varepsilon}
\newcommand{\al}{\alpha}

\newcommand{\ga}{\gamma}
\newcommand{\G}{\Gamma}
\newcommand{\de}{\delta}
\newcommand{\De}{\Delta}

\newcommand{\Si}{\Sigma}

\newcommand{\ka}{\kappa}
\newcommand{\ta}{\tau}

\newcommand{\pd}{\partial}

\newcommand{\co}{{\cal O}}

\newcommand{\ov}[1]{\overline{#1}}

\newcommand{\ev}[1]{<\!\!{#1}\!\!>}

\title{Quarks and gluons in a magnetic field}
\ShortTitle{Quarks and gluons in a magnetic field}
\author{\speaker{Peter Watson}%
         \thanks{Work supported by the Deutsche Forschungsgemeinschaft 
(DFG) under contract no. DFG-Re856-9/1.}\\
Institut f\"ur Theoretische Physik, Universit\"at T\"ubingen, 
Auf der Morgenstelle 14, 72076 T\"ubingen, Deutschland\\

        E-mail: \email{peter.watson@uni-tuebingen.de}}
\author{Hugo Reinhardt\\
Institut f\"ur Theoretische Physik, Universit\"at T\"ubingen, 
Auf der Morgenstelle 14, 72076 T\"ubingen, Deutschland
}

\abstract{
The quark gap equation under the rainbow truncation, with two versions of a phenomenological one-gluon exchange interaction and in the presence of a uniform magnetic field is considered.  It is argued that in order to describe the quark condensate in the limit of vanishing magnetic fields, one must sum over the Landau levels.  The resulting chiral quark condensate rises quadratically for small magnetic fields and linearly for large fields, in qualitative agreement with various recent lattice results.  It is observed that when discussing quarks, the magnitude of the magnetic field must be considered relative to the scale of the strong interaction.
}

\FullConference{From quarks and gluons to hadronic matter: A bridge too far?\\
                 2-6 September, 2013\\
                 European Centre for Theoretical Studies in Nuclear Physics and Related Areas (ECT*), Villazzano, Trento (Italy)}
\begin{document}
\section{Introduction}

In this talk, we shall discuss the problem of quarks in an external magnetic field\footnote{The talk is based on work done in the early stages of writing Ref.~\cite{Watson:2013ghq}: for a full account of the material and results (including further results obtained after the workshop had taken place), we refer the reader to the article.}.  The topic of fermions in the presence of a magnetic field, especially a strong magnetic field, has attracted considerable attention recently (see Ref.~\cite{Shovkovy:2012zn} for a recent review) and the primary focus is on magnetic catalysis -- the observation that the fermion condensate is increased by the magnetic field.  Chiral symmetry breaking in the presence of a magnetic field has, however, been studied for quite some time: early studies looked at, for example, the Gross-Neveu model \cite{Klimenko:1990rh,Klimenko:1991he,Klimenko:1992ch} and quantum electrodynamics (QED) \cite{Gusynin:1994xp,Gusynin:1995gt}.  The solution to the Dirac equation with an external magnetic field leads to the appearance of Landau levels and this forms the basis of the so-called Ritus eigenfunction method \cite{Ritus:1978cj,Ritus:proc} for describing the problem.  It is argued that in QED with large magnetic fields, the lowest Landau level dominates the chiral symmetry breaking \cite{Gusynin:1994xp,Gusynin:1995gt}.  Here, we consider the case of quantum chromodynamics (QCD), i.e., quarks.  In contrast to the electrons in QED (at least for physical values of the coupling), the quarks experience dynamical chiral symmetry breaking even in the absence of the magnetic field by virtue of the strong interaction.  As will be shortly demonstrated, the expansion in terms of Landau levels is not directly applicable in this context and a summation is required -- it is necessary to consider the case of small magnetic fields.  However, for quarks, the strength of the magnetic field should be judged relative to the scale of the strong interaction and what is described by the `small' field limit turns out to include the estimated fields present in noncentral heavy-ion collisions \cite{Skokov:2009qp} (which are usually thought of as being very large).

\section{Ritus eigenfunction method}
Throughout these proceedings, we shall use the notation and conventions of Ref.~\cite{Watson:2013ghq}.  We consider quarks in the presence of a uniform magnetic field, $\vec{B}=B\hat{e}_3$, and for simplicity, restrict to the case $h\equiv QB\geq0$, where $Q$ is the electromagnetic charge of the quark in question.  The electromagnetic gauge potential associated with the magnetic field is chosen as $A^0=0$, $\vec{A}=Bx_1\hat{e}_2$ and is minimally coupled in the quark component of the action to give the Dirac operator
\be
D-m=\imath\pd_{\mu}\ga^{\,\mu}-h\ga^{\,2}x_1-m.
\ee
The tree-level proper two-point function, $\G^{(0)}$, is defined by
\be
\G^{(0)}(x,y)=\imath[D-m]\de(x-y),
\ee
whereas the corresponding tree-level propagator, $S^{(0)}$, is the solution of
\be
\imath[D-m]S^{(0)}(x,y)=\de(x-y).
\ee
As is well-understood from quantum mechanics, the presence of a uniform magnetic field in the Dirac equation leads to the introduction of Landau levels, with Hermite functions as eigenfunctions.  The various Landau levels are intimately connected to the spin states.  Further, the presence of the electromagnetic gauge potential is such that translational invariance is broken.  The problem of fermions in a magnetic field was studied extensively by Ritus using eigenfunction methods \cite{Ritus:1978cj,Ritus:proc}.  The tree-level proper two-point function (the expression for the propagator is similar) can be written (in our notation \cite{Watson:2013ghq}) as
\be
\G^{(0)}(x,y)=\sum_{n=0}^{\infty}\int\frac{d^3\tilde{p}}{(2\pi)^3}E(x;\tilde{p},n)\G^{(0)}(\ov{p},n)\ov{E}(y;\tilde{p},n),
\label{eq:pinvdec}
\ee
where $\tilde{p}^{\mu}=(p_0,0,p_2,p_3)$, $\ov{p}^{\mu}=(p_0,0,0,p_3)$, and the discrete index $n=0,1,\ldots$ labels the Landau levels.  ($\ov{E}$) $E$ is the so-called (conjugate) Ritus matrix, given by
\be
E(x;\tilde{p},n)=h^{1/4}e^{-\imath\tilde{p}\cdot x}\left[\psi_{n-1}(\e)\Si^++\psi_{n}(\e)\Si^-\right].
\label{eq:rituse}
\ee
In the above, the $\psi_n$ are Hermite functions with argument $\e=\sqrt{h}x_1+p_2/\sqrt{h}$.  The factors $\Si^{\pm}=\left[1\pm\imath\ga^{\,1}\ga^{\,2}\right]/2$ are spin projection operators.  It can be shown that the Ritus matrices, $E$ and $\ov{E}$, are orthonormal and form a complete set.  The idea is to use these matrices as a replacement for the usual Fourier exponential factors $e^{-\imath p\cdot x}$.  The eigenvalues $\ov{p}$ and $n$ then replace the standard momentum space components $p^{\mu}$.  In this space, the tree-level proper two-point function is given by
\be
-\imath\G^{(0)}(\ov{p},n)=\ov{p}_{\mu}\ga^{\,\mu}-\sqrt{2nh}\ga^{\,2}-m.
\ee
In similar fashion, the tree-level propagator is given by
\be
\imath S^{(0)}(\ov{p},n)=\frac{\ov{p}_{\mu}\ga^{\,\mu}-\sqrt{2nh}\ga^{\,2}+m}{\ov{p}^2-2nh-m^2+\imath0_+}.
\ee

Let us now demonstrate why the Ritus method cannot be directly applied to the problem of quarks in a magnetic field.  With this approach, the dressed propagator has the general form \cite{Watson:2013ghq}
\be
S(x,y)=\sum_{n=0}^{\infty}\int\frac{d^3\tilde{p}}{(2\pi)^3}E(x;\tilde{p},n)S(\ov{p},n)\ov{E}(y;\tilde{p},n).
\label{eq:propgen}
\ee
In the above, it is assumed that the quark-gluon interaction of the self-energy is not affected by the presence of the magnetic field (the gluon has no electric charge, so this will be true at leading order), such that the general form of the propagator has the same form as its tree-level counterpart.  Using the properties of the Ritus matrices, the quark condensate is given by (trace over Dirac matrices)
\be
\ev{\ov{q}q}=N_c\mbox{Tr}_dS(x,x)=N_c\frac{h}{2\pi}\int\frac{d^2\ov{p}}{(2\pi)^2}\mbox{Tr}_d\left\{\Si^-S(\ov{p},n=0)+\sum_{n=1}^{\infty}S(\ov{p},n)\right\}.
\ee
The important point here is the appearance of the pre-factor $h$ (the factor $h$ also appears in all loop integrals).  It arises quite generally because the dressed propagator in the eigenvalue space is a function of (the two components of) $\ov{p}$ and discrete index $n$ instead of four-momentum $p^{\,\mu}$ -- the only scale that can compensate for the change in the dimensions of the integral is $h$.    Naively then, in the limit $h\rightarrow0$, the condensate (and the self-energy) would vanish.  This is clearly not true -- we know full well that chiral symmetry is dynamically broken for quarks at zero temperature.  The resolution to this lies in the recognition that the expansion of the two-point functions in terms of Landau levels has the characteristic behavior of an asymptotic series.  To get the correct $h\rightarrow0$ limit, we must first sum the series.
\section{Summation and Schwinger phase}
In the case of the tree-level two-point functions, it is known how to sum over the Landau levels \cite{Gorbar:2013upa,Gorbar:2013uga}.  Let us illustrate this for the proper tree-level two-point function.  In the expression for $\G^{(0)}$, \eq{eq:pinvdec}, there arises the following type of integral ($\ta=\sqrt{h}y_1+p_2/\sqrt{h}$, $a$ and $b$ are indices)
\be
I_{ab}=\int dp_2\,e^{\imath p_2(x_2-y_2)}\psi_{a}(\e)\psi_{b}(\ta).
\label{eq:intiab}
\ee
It can be shown that such integrals can \emph{almost} be written in terms of a Fourier integral over transverse momentum components, $\vec{p}_t=(p_1,p_2,0)$ and Laguerre polynomials ($L_m^\al$): schematically (see Ref.~\cite{Watson:2013ghq} for the full expressions)
\be
I_{ab}\sim e^{\imath\Phi}\int d^2p_t\,e^{\imath\vec{p}_t\cdot(\vec{x}-\vec{y})}f(\vec{p}_t)\exp{\left\{-\frac{p_t^2}{h}\right\}}L_m^\al\left(\frac{2p_t^2}{h}\right)
\ee
where $f(\vec{p}_t)$ is a collection of various factors and $m,\al$ are combinations of the original indices $a$ and $b$.  The ``\emph{almost}'' in the previous sentence refers to the appearance of the factor $\Phi$, the Schwinger phase, which reads here
\be
\Phi=-\frac{h}{2}(x_2-y_2)(x_1+y_1).
\ee
This factor encodes the deviations from translational invariance inherent to the presence of the magnetic field.  Expanding \eq{eq:pinvdec} in terms of the above integrals $I_{ab}$ and evaluating in terms of the Laguerre polynomials, the tree-level proper two-point function then reads \cite{Watson:2013ghq,Gorbar:2013upa,Gorbar:2013uga}
\bea
\lefteqn{-\imath\G^{(0)}(x,y)=e^{\imath\Phi}\int\frac{d^4p}{(2\pi)^4}e^{-\imath p\cdot(x-y)}e^{-p_t^2/h}}&&\nonumber\\
&&\times\sum_{n=0}^{\infty}2(-1)^{n}\left\{\left[\Si^-L_n\left(\frac{2p_t^2}{h}\right)-\Si^+L_{n-1}\left(\frac{2p_t^2}{h}\right)\right](\ov{p}_{\mu}\ga^{\,\mu}-m)+2L_{n-1}^{1}\left(\frac{2p_t^2}{h}\right)\vec{p}_t\cdot\vec{\ga}\right\}.
\label{eq:invnnn}
\eea
The sums over $n$ are known and one gets the final result
\be
-\imath\G^{(0)}(x,y)=e^{\imath\Phi}\int\frac{d^4p}{(2\pi)^4}e^{-\imath p\cdot(x-y)}\left[p_{\mu}\ga^{\,\mu}-m\right].
\ee
The only effect of the magnetic field on the tree-level proper two-point function is the introduction of the Schwinger phase.  The tree-level propagator can be treated in similar fashion (using further results obtained originally in Ref.~\cite{Chodos:1990vv}) and for small $h$ the expression reads \cite{Watson:2013ghq,Gorbar:2013upa,Gorbar:2013uga}
\be
\imath S^{(0)}(x,y)=e^{\imath\Phi}\int\frac{d^4p}{(2\pi)^4}e^{-\imath p\cdot(x-y)}\left\{\frac{p_{\mu}\ga^{\,\mu}+m}{p^2-m^2+\imath0_+}+\frac{\imath h\ga^{\,1}\ga^{\,2}[\ov{p}_{\mu}\ga^{\,\mu}+m]}{[p^2-m^2+\imath0_+]^2}+\co(h^2)\right\}.
\label{eq:proptree}
\ee
One can see that after summation, the standard tree-level two-point functions emerge as $h\rightarrow0$ (unlike for the Ritus expanded expressions in the last section).  However, it is not obvious that one is the inverse of the other for nonzero $h$.

\section{Nonperturbative propagator}
In order to solve the quark gap equation, we need to know the connection between the nonperturbatively dressed proper two-point function and the propagator.  To do this, we take a step back to the Ritus decomposition in terms of Landau levels.  The proper two-point function has the general form \cite{Watson:2013ghq}
\be
\G(x,y)=\sum_{n=0}^{\infty}\int\frac{d^3\tilde{p}}{(2\pi)^3}E(x;\tilde{p},n)\G(\ov{p},n)\ov{E}(y;\tilde{p},n)
\label{eq:invgen}
\ee
(as for the propagator, \eq{eq:propgen}, we assume that the magnetic field does not modify the quark-gluon interaction) and we consider the following decomposition in eigenvalue space
\be
-\imath\G(\ov{p},n)=\Si^+(\ov{p}_{\mu}\ga^{\,\mu}A-B)+\Si^-(\ov{p}_{\mu}\ga^{\,\mu}C-D)-\sqrt{2nh}\ga^{\,2}E+\sqrt{2nh}[\Si^+-\Si^-]\ov{p}_{\mu}\ga^{\,\mu}\ga^{\,2}F,
\ee
where the dressing functions, $A$-$F$ are functions of $\ov{p}^2$ and $n$ (arguments being suppressed for brevity).  This decomposition allows for different spin contributions with the projectors $\Si^\pm$ inherent to the presence of the magnetic field.  With the general form, \eq{eq:propgen}, and using the orthonormality and completeness properties of the Ritus matrices, it is possible to find the corresponding expression for the propagator in terms of the Ritus decomposition.  The expression looks like (writing only the first term and neglecting the Feynman prescription)
\be
\imath S(\ov{p},n)=\Si^+\ov{p}_{\mu}\ga^{\,\mu}\frac{\De_1C-\De_2D}{\De}+\ldots,
\label{eq:propp}
\ee
where
\be
\De_1=\ov{p}^2AC-BD-2nhE^2-2nh\ov{p}^2F^2,\;\;\De_2=4nhEF+AD-BC,\;\;\De=\De_1^2-\ov{p}^2\De_2^2
\label{eq:propden}
\ee
(again, see \cite{Watson:2013ghq} for the full expressions).

We now have to make some approximations to sum over the Landau levels.  The first approximation is to drop the dressing function $F$ (and the corresponding component in the propagator); this is justified on the grounds that the accompanying Dirac structure is not present at tree-level or in the absence of the magnetic field.  Given the form, \eq{eq:invgen}, for the proper two-point function, the integrals $I_{ab}$, \eq{eq:intiab}, can be extracted as for the tree-level case to give an expression similar to \eq{eq:invnnn} (this also applies for the propagator).  The sum over the Landau levels can be done and gives rise to a new set of dressing functions $\hat{A}$-$\hat{E}$ with arguments $\ov{p}^2$ and $p_t^2$.  The explicit expression for the dressed proper two-point function is \cite{Watson:2013ghq}
\be
-\imath\G(x,y)=e^{\imath\Phi}\int\frac{d^4p}{(2\pi)^4}e^{-\imath p\cdot(x-y)}\left[\Si^+(\ov{p}_{\mu}\ga^{\,\mu}\hat{A}-\hat{B})+\Si^-(\ov{p}_{\mu}\ga^{\,\mu}\hat{C}-\hat{D})-\vec{p}_t\cdot\vec{\ga}\,\hat{E}\right].
\label{eq:invp}
\ee
At tree-level $\hat{A}=\hat{C}=\hat{E}=1$, $\hat{B}=\hat{D}=m$.  In the absence of the magnetic field ($h=0$), we should see (this will be verified numerically) that the proper two-point function reduces to its standard covariant form in momentum space: $-\imath\G(p)=p_{\mu}\ga^{\,\mu}A_L-B_L$, with two dressing functions $A_L$ and $B_L$ (both being functions of $p^2$).

The propagator, \eq{eq:propp} contains nontrivial denominator factors, \eq{eq:propden}, including dressing functions whose dependence on the Landau levels, $n$, is not \emph{a priori} known.  To perform the sum, we now assume that the dressing functions $A$-$E$ may be replaced by their counterparts $\hat{A}$-$\hat{E}$ before summing.  Further, we expand the denominator factors around $h=0$.  This leads to the following expression for the propagator at small $h$ \cite{Watson:2013ghq}
\bea
\lefteqn{\imath S(x,y)=e^{\imath\Phi(x,y)}\int\frac{d^4p}{(2\pi)^4}
e^{-\imath p\cdot(x-y)}}&&\nonumber\\
&&\times\left\{
\frac{1}
{[\ov{p}^2\hat{A}\hat{C}-p_t^2\hat{E}^2-\hat{B}\hat{D}+\imath0_+]}
\left[\Si^+(\ov{p}_{\mu}\ga^{\,\mu}\hat{C}+\hat{D})
+\Si^-(\ov{p}_{\mu}\ga^{\,\mu}\hat{A}+\hat{B})
-\vec{p}_t\cdot\vec{\ga}\,\hat{E}\right]
\right.\nonumber\\&&\left.
+\frac{h\hat{E}^2}
{[\ov{p}^2\hat{A}\hat{C}-p_t^2\hat{E}^2-\hat{B}\hat{D}+\imath0_+]^2}
\left[\Si^+(\ov{p}_{\mu}\ga^{\,\mu}\hat{C}+\hat{D})
-\Si^-(\ov{p}_{\mu}\ga^{\,\mu}\hat{A}+\hat{B})\right]
\right.\nonumber\\&&\left.
+\frac{(\hat{A}\hat{D}-\hat{B}\hat{C})}
{[\ov{p}^2\hat{A}\hat{C}-p_t^2\hat{E}^2-\hat{B}\hat{D}+\imath0_+]^2}
\left[-\Si^+(\ov{p}_{\mu}\ga^{\,\mu}\hat{D}+\ov{p}^2\hat{C})
+\Si^-(\ov{p}_{\mu}\ga^{\,\mu}\hat{B}+\ov{p}^2\hat{A})\right]
\right\}.
\label{eq:prop}
\eea
The above expression reduces explicitly to its tree-level counterpart, \eq{eq:proptree}, and to the standard covariant gauge form in the absence of the magnetic field.  The approximations are such that the connection between the various Dirac structures that appear in the two-point functions due to the presence of the magnetic field is maintained.  To justify the replacement of the $n$-dependent dressing functions $A$-$E$ with the $p_t^2$-dependent functions $\hat{A}$-$\hat{E}$, we notice that for dimensional reasons, the occurrence of $n$ within the dressing functions $A$-$E$ would typically be associated with a factor $h$ (consider, for example, the tree-level propagator denominator factor: $\ov{p}^2-2nh-m^2$) such that any $n$-dependence would be suppressed for small values of $h$.  Moreover, it is the dynamics of the gap equation that will decide on the eventual $p_t^2$-dependence of the dressing functions $\hat{A}$-$\hat{E}$.

With \eq{eq:prop} for the propagator, the explicit expression for the condensate is given by
\begin{align}
&\ev{\ov{q}q}=N_c\mbox{Tr}_dS(x,x)\nonumber\\
&=-2\imath N_c\int\frac{d^4p}{(2\pi)^4}\left\{\frac{\hat{B}+\hat{D}}{[\ov{p}^2\hat{A}\hat{C}-p_t^2\hat{E}^2-\hat{B}\hat{D}+\imath0_+]}+\frac{h\hat{E}^2(\hat{D}-\hat{B})+\ov{p}^2(\hat{A}\hat{D}-\hat{B}\hat{C})(\hat{A}-\hat{C})}{[\ov{p}^2\hat{A}\hat{C}-p_t^2\hat{E}^2-\hat{B}\hat{D}+\imath0_+]^2}\right\}.
\end{align}

\section{Gap equation}
To calculate the nonperturbative quark propagator, we solve the rainbow truncated gap (Dyson-Schwinger) equation.  In configuration space and after resolving the color factors ($C_F=4/3$ for $N_c=3$ colors), this reads
\be
\G(x,y)=\G^{(0)}(x,y)+g^2C_F\ga^{\,\mu}S(x,y)\ga^{\,\ka}W_{\ka\mu}(y,x).
\label{eq:gap}
\ee
We use a dressed gluon interaction of the (Landau gauge) form
\be
\imath W_{\ka\mu}(y,x)=\int\frac{d^4q}{(2\pi)^4}e^{-\imath q\cdot(y-x)}\frac{G(q^2)}{q^2}t_{\ka\mu}(q)
\ee
where $t_{\ka\mu}(q)=g_{\ka\mu}-q_{\mu}q_{\mu}/q^2$ is the transverse momentum projector.  For the dressing function $G$, we take either of the two phenomenological forms (in Minkowski space)
\be
g^2\frac{G(q^2)}{q^2}=4\pi^2d\exp{\left\{\frac{q^2}{\w^2}\right\}}\times\left\{\begin{array}{cc}q^2/\w^2,&\mbox{I}\\-1,&\mbox{II}\end{array}\right..
\ee
The first of these forms (type I) was used previously \cite{Alkofer:2002bp} to study dynamical chiral symmetry breaking and light meson phenomenology: we shall use the parameters $\w=0.5\,\mbox{GeV}$, $d=16\,\mbox{GeV}^{-2}$.  The second form (type II) is a variation of this and is inspired by kernels constructed from lattice components \cite{Aguilar:2010cn}: for this interaction we use $\w=0.5\,\mbox{GeV}$, $d=41\,\mbox{GeV}^{-2}$ (these parameters chosen such that the chiral condensate in the absence of the magnetic field is roughly the same for the two interaction types).  Inserting the nonperturbative forms for the quark propagator, \eq{eq:prop}, and proper two-point function, \eq{eq:invp}, into \eq{eq:gap}, the Schwinger phase cancels.  After Wick rotating to Euclidean space ($p_0\rightarrow\imath p_4$), one can numerically solve for the various dressing functions (as functions of two variables, $p_l^2=p_3^2+p_4^2$ and $p_t^2=p_1^2+p_2^2$).  We will consider only chiral quarks.

\section{Results}

\begin{figure}[t]
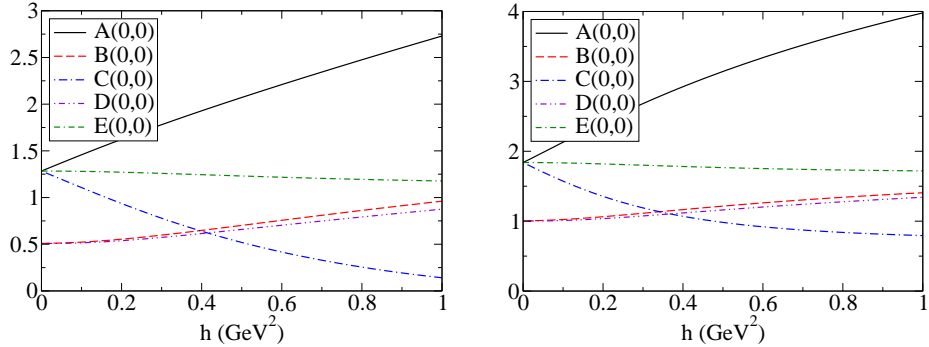

\vspace{0.8cm}
\begin{center}
\includegraphics[width=0.385\linewidth]{dresshplot.eps}\hspace{0.8cm}\includegraphics[width=0.37\linewidth]{dresshplot2.eps}
\end{center}
\caption{\label{fig:dressh}Plot of the dressing functions $\hat{A}$-$\hat{E}$ evaluated at $p_l^2=p_t^2=0$ as a function of $h$ for the type I (left panel) and type II (right panel) interactions.  See text for details.}
\end{figure}

In Fig.~\ref{fig:dressh}, we illustrate the effect of the magnetic field on the various dressing function for both types of interaction in the case of chiral quarks.  The functions $\hat{A}$-$\hat{E}$ evaluated at $p_l^2=p_t^2=0$ are plotted as functions of $h$.  What is clearly seen is the reduction (for both interaction types) of the functions $\hat{A}$-$\hat{E}$ to the standard covariant gauge case with $A_L$ and $B_L$ as $h\rightarrow0$.  It can be explicitly verified that when $h=0$, the functions for the type I interaction match those obtained in Ref.~\cite{Alkofer:2002bp} (from which the interaction and parameters were taken), confirming that the $h\rightarrow0$ limit is respected in the approximations.  Further, it is seen that the qualitative pattern of results is the same for both types of interaction.  This is to be expected because at this level of truncation, the interaction is not modified by the presence of the magnetic field.

\begin{figure}[t]
\vspace{0.8cm}
\begin{center}
\includegraphics[width=0.4\linewidth]{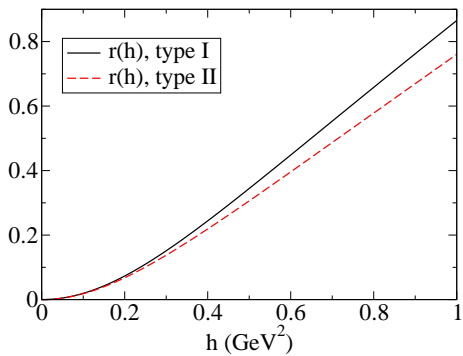}
\end{center}
\caption{\label{fig:rh}Plot of the (dimensionless) ratio $r(h)$, as a function of $h$ for type I and type II interactions.  See text for details.}
\end{figure}

In the absence of the magnetic field and with the parameter sets given in the previous section, the chiral quark condensate is
\be
\ev{\ov{q}q}_{h=0}=(-251\,\mbox{MeV})^3
\ee
(for both interaction types).  The change in the chiral condensate due to the magnetic field is conventionally expressed in terms of the (dimensionless) relative increment function, $r(h)$, defined as
\be
r(h)=\frac{\ev{\ov{q}q}_h}{\ev{\ov{q}q}_{h=0}}-1.
\label{eq:rhdef}
\ee
This is plotted in Fig.~\ref{fig:rh} for both interaction types.  It is seen that the condensate increases quadratically with $h$ for small $h$ and linearly for large $h$ (the transition occurs at around $h\approx0.3\,\mbox{GeV}^2$).  Qualitatively, this agrees with recent lattice results, e.g., Refs.~\cite{D'Elia:2011zu,D'Elia:2013twa,Bali:2012zg,Bali:2013cf,Buividovich:2008wf} (see also \cite{Simonov:2012mf}).  However, we should point out that a linear rise for small $h$ is seen in Ref.~\cite{Ilgenfritz:2012fw} (see also M.~M\"uller-Preussker's contribution to these proceedings) and for chiral perturbation theory \cite{Shushpanov:1997sf}.  In Ref.~\cite{Watson:2013ghq}, a more detailed comparison to the lattice data of Ref.~\cite{D'Elia:2011zu} was performed: it was seen that the small $h$ quadratic behavior and the position of the transition between the quadratic and linear regimes is well-reproduced.  The position of the transition ($h\approx0.3\,\mbox{GeV}^2$) is rather important because this value coincides with estimates for the maximum magnitude of the magnetic fields expected in noncentral heavy-ion collisions \cite{Skokov:2009qp}.  It would thus appear that while the magnetic fields may be large in such systems, when discussing quarks, such fields are competing with the scales associated with the strong interaction and it is the `small' $h$ behavior of the condensate that is relevant, not the large $h$ behavior.

\begin{figure}[t]
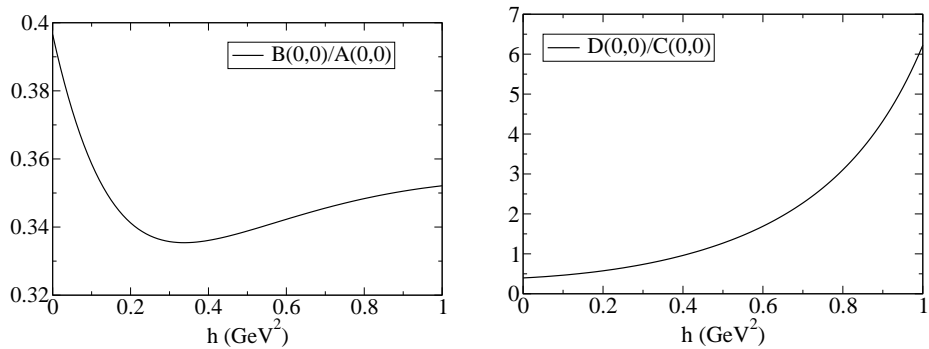

\vspace{0.8cm}
\begin{center}
\includegraphics[width=0.385\linewidth]{lhplot1.eps}\hspace{0.8cm}\includegraphics[width=0.37\linewidth]{lhplot2.eps}
\end{center}
\caption{\label{fig:lhplot}Plot of the ratios of dressing functions $\hat{B}/\hat{A}$ (left panel) and $\hat{D}/\hat{C}$ (right panel) evaluated at $p_l^2=p_t^2=0$ as a function of $h$ for the type I interaction.  See text for details.}
\end{figure}

It has long been argued (see, e.g., Refs.~\cite{Gusynin:1994xp,Gusynin:1995gt}) that in the fermion gap equation for large $h$ (which in the context here means larger than the scale of the strong interaction), the lowest ($n=0$) Landau level of the Ritus decomposition is dominant, such that the system may be described in the so-called lowest Landau level approximation.  In this approximation, the characteristic structure of the Ritus matrix $E$, \eq{eq:rituse}, reduces
\be
[\Si^+\psi_{n-1}(\e)+\Si^-\psi_{n}(\e)]\stackrel{n=0}{\longrightarrow}\psi_{0}(\e)\Si^-.
\ee
This has the effect of projecting the quark propagator onto the $\Si^-$ spin structures.  Having summed over the Landau levels to obtain the nonperturbative two-point functions (with approximations tailored to studying the small $h$ limit), we have no direct access to the individual Landau level contributions.  However, we do see that the $\Si^-$ spin structure is dominant.  The dressing functions $\hat{C}$ and $\hat{D}$ correspond to the $\Si^-$ spin components in the decomposition \eq{eq:invp}, whereas $\hat{A}$ and $\hat{B}$ correspond to the $\Si^+$ components.  The ratios (analogue to the standard quark mass function) $\hat{B}/\hat{A}$ and $\hat{D}/\hat{C}$ for the type I interaction and evaluated at $p_l^2=p_t^2=0$ are plotted as a function of $h$ in Fig.~\ref{fig:lhplot}.  It is seen that the ratio $\hat{D}/\hat{C}$ increases dramatically (for the type II interaction, this increase is not quite so big) whereas the ratio $\hat{B}/\hat{A}$ is comparatively stable.  The mechanism behind this is the decrease of the function $\hat{C}$ with increasing $h$ (see Fig.~\ref{fig:dressh}).  Thus, even with our approximations tailored to the $h\rightarrow0$ limit, we see that the spin structures associated with the lowest Landau level are indeed dominant (at least qualitatively).

\section{Summary}
We have looked at the chiral quark gap equation under the rainbow truncation, with a phenomenological gluon interaction and in the presence of a constant magnetic field \cite{Watson:2013ghq}.  In the presence of a magnetic field, the problem can be posed in terms of the Ritus eigenfunction method \cite{Ritus:1978cj} (expanding in terms of Landau levels).  However, for small magnetic fields relative to the scale of the strong interaction, such an expansion results in a vanishing quark condensate.  To overcome this, we must sum over the Landau levels.  This was done using various approximations and the gap equation was solved.  It is seen that the quark two-point functions and condensate reduce to their standard covariant gauge forms in the absence of the magnetic field.  The condensate rises quadratically for small magnetic fields and linearly for large fields, in qualitative agreement with recent lattice results \cite{D'Elia:2011zu,D'Elia:2013twa,Bali:2012zg,Bali:2013cf,Buividovich:2008wf}.  The transition region between the two types of behavior lies at roughly the scale associated with the maximum fields estimated to be present in noncentral heavy-ion collisions \cite{Skokov:2009qp}, such that in the context of QCD, the magnitude of the magnetic field should be interpreted relative to the scale of the strong interaction.

There are two interesting avenues for future work.  Firstly, the case of finite temperature and chemical potential might be investigated in order to study the phase diagram in the presence of the magnetic field.  Second, unquenching effects should be considered (in this respect, the inclusion of quark loops to the phenomenological interactions used here was studied previously in Ref.~\cite{Fischer:2005en}).

\begin{acknowledgments}
It is a pleasure to thank the organizers (D.~Binosi in particular) for a most enjoyable and informative workshop.
\end{acknowledgments}

\end{document}